\documentclass[letterpaper,10pt,conference]{IEEEtran}

\usepackage{amssymb}
\usepackage[figuresright]{rotating}
\usepackage{amsmath,graphicx,epsfig,latexsym,mathrsfs,epsf,bm}
\usepackage{amsthm}
\usepackage{color}
\usepackage{epstopdf}
\def\Pr{\mathop{\textrm{Pr}}}

\newsavebox{\fminibox}
\newlength{\fminilength}

  \def\+{^\dagger}

\def\nequiv{\not\kern-.05em\equiv}
\def\egal{\kern-.5em=\kern-.5em}        
\def\propt{\kern-.2em\propto\kern-.2em} 

\def\intdouble{\int\kern-0.3em\int}
\def\inttriple{\int\kern-0.3em\int\kern-0.3em\int}

\def\rond#1{\overset{\kern-0.33em~_\circ}{#1}}
\def\rondit[#1]#2{\overset{\kern#1~_\circ}{#2}}

\usepackage{subcaption}
\usepackage{mathabx}
\usepackage{xfrac}
\usepackage{upgreek}
\usepackage{verbatim}
\usepackage{xparse}
\NewDocumentCommand{\ceil}{s O{} m}{%
  \IfBooleanTF{#1} 
    {\left\lceil#3\right\rceil} 
    {#2\lceil#3#2\rceil} 
}

\newtheorem{Theorem}{Theorem}

\newtheorem{CounterExample*}{$\overline{\hbox{\bf Example}}$}
\newtheorem{Definition}{Definition}

\newtheorem{Example*}{Example}

\newtheorem{Intuition*}{Intuition}
\newtheorem{Joke*}{Joke}
\newtheorem{Lemma}[Theorem]{Lemma}
\newtheorem{Lemma*}{Lemma}

\newtheorem{Note*}{Note}
\newtheorem{Open problem}{Open problem}
\newtheorem{Proposition}{Proposition}

\newtheorem{Question*}{Question}
\newtheorem{Remark*}{Remark}

\newcommand{\uXM}{\underline{{X}}_{\mathcal{M}}}

\newcommand{\tPr}{\textnormal{Pr}}

\newcommand{\exc}{{\rm  E}}
\newcommand{\mcN}{{\mathcal N}}
\newcommand{\mcM}{{\mathcal M}}
\newcommand{\mcE}{{\mathcal E}}

\newcommand{\mcH}{{\mathcal H}}

\newcommand{\mcV}{{\mathcal V}}
\newcommand{\mcD}{{\mathcal D}}
\newcommand{\mcG}{{\mathcal G}}
\newcommand{\mcX}{{\mathcal X}}

\newcommand{\mbbR}{{\mathbb R}}

\newcommand{\uzm}{\underline{0}}
\newcommand{\ux}{\underline{{x}}}

\newcommand{\urho}{\underline{\rho}}
\newcommand{\uv}{\underline{v}}

\newcommand{\uW}{\underline{W}}
\newcommand{\uZ}{\underline{Z}}

\newcommand{\uX}{\underline{{X}}}

\newcommand{\bSigma}{\mathbf{\Sigma}}

\newcommand{\bDelta}{\mathbf{\Delta}}

\newcommand{\bI}{{\bf I}}

\newcommand{\bK}{{\bf K}}

\author{\IEEEauthorblockN{Navid Tafaghodi Khajavi and Anthony Kuh}
\IEEEauthorblockA{Department of Electrical  Engineering,\\
University of Hawaii, Honolulu, HI 96822\\
Email: navidt@hawaii$.$edu, kuh@hawaii$.$edu}}

\begin{document}

\title{The Goodness of Covariance Selection Problem from AUC Bounds}


\maketitle

\begin{abstract}


We conduct a study of graphical models and discuss the quality of model selection approximation by formulating the problem as a detection problem and examining the area under the curve (AUC).
We are specifically looking at the model selection problem for jointly Gaussian random vectors. 
For Gaussian random vectors, this problem simplifies to the covariance selection problem which is widely discussed in literature by Dempster \cite{dempster}.
In this paper, we give the definition for the correlation approximation matrix (CAM) which contains all information about the model selection problem and discuss the $p$th order Markov chain model and the $p$th order star network model for the a Gaussian distribution with Toeplitz covariance matrix. For each model, we compute the model covariance matrix as well as the KL divergence between the Gaussian distribution and its model. 
We also show that if the model order, $p$, is proportional to the number of nodes, $n$, then the model selection is asymptotically good as the number of nodes, $n$, goes to infinity since the AUC in this case is bounded away from one.
We conduct some simulations which confirm the theoretical analysis and also show that the selected model quality increases as the model order, $p$, increases.

\end{abstract}

\section{Introduction}
\label{sec:intro}

In signal processing and machine learning a fundamental problem is to balance performance quality (i.e. minimizing cost function) with computational complexity.
A powerful tool in order to address this trade-off is graphical model selection. Model selection methods provide approximated models with desired accuracy as needed for different applications. 
Given data, different model selection algorithms impose different structure to model data. 
In the case of jointly Gaussian data, covariance selection problem is presented and studied in \cite{dempster} and \cite{GMbook96}.
The purpose of the covariance selection problem is to reduce the computation complexity in various applications.

Some of the model selection algorithms to impose structure are the Chow-Liu minimum spanning tree (MST) \cite{chowliu}, the first order Markov chain approximation \cite{ISIT2013} and penalized likelihood methods such as LASSO \cite{LASSO} and graphical LASSO \cite{GLASSO} that can be used to approximate the correlation matrix and inverse correlation matrix with a more sparse graph while retaining good accuracy. 
The Chow-Liu MST algorithm for Gaussian distribution is to find the optimal tree structure using a Kullback-Leibler (KL) divergence cost function \cite{dempster}. The Chow-Liu algorithm utilizes the Kruskal algorithm \cite{kruskal}.
The first order Markov chain approximation uses a regret cost function to output a chain structured graph \cite{ISIT2013}. 
Penalized likelihood methods specify the graph representation by eliminating some of the edges.

In this paper we extend work of \cite{MQ16ArXiv} where we formulated a covariance model selection paper using a detection problem formulation.  The \cite{MQ16ArXiv} focused on examples where approximation were trees.  Here we extend approximations to clique graphs with junction trees. We consider a simple example where the covariance matrix is a Toeplitz covariance matrix with ones along the diagonal and correlation coefficient $\rho$ on the off-diagonals.
This covariance matrix is interesting and arise in different applications\footnote{Looking at the solar irradiation datasets \cite{APSIPA2014}, we can see that sensors that are distributed in small geographical areas are highly correlated and have approximately the same correlations.}.
Given this covariance matrix, we ask the following question, "when is a covariance selection approximation good?"
To answer this question we use the detection problem formulation proposed in \cite{MQ16ArXiv}.
The detection problem for Gaussian data leads to calculation of the log-likelihood ratio test (LLRT), the receiver operating characteristic (ROC) curve, the KL divergence and the reverse KL divergence as well as the area under the curve (AUC) where the AUC is used as the accuracy measure for the detection problem on average.
We also present the correlation approximation matrix (CAM) as the product of the original correlation matrix and the inverse of the model approximation correlation matrix. For Gaussian data this matrix contains all the information needed to compute the information divergences, the ROC curve and the area under this curve, i.e. the AUC.
We present an analytical expression to compute the KL divergence between the original distribution and the model covariance matrix of order, $p$.
We show that if we pick a model order, $p$, proportional to the number of nodes, $n$, the AUC is asymptotically bounded away from one as $n$ goes to infinity.
Moreover, we present some simulation results. We pick different values as the order of the approximation model and compare the $p$th order Star approximation model with the $p$th order Markov chain approximation model.
Simulation results show that the $p$th order star approximation model has smaller AUC than the $p$th order Markov chain approximation model and thus has better performance.
Also, through simulations we confirm our theoretical results showing that the AUC is bounded away from one when model order, $p$, is proportional to the number of nodes, $n$.

The rest of this paper is organized as follows. 
In section \ref{sec:prob} we give the detection problem framework, the sufficient test statistic and the log-likelihood ratio test. Moreover, the sufficient test statistic for Gaussian data as well as its distribution under both hypotheses are also presented in this section.
The ROC curve and the definition of AUC as well as analytical expression for the AUC are presented in this section.
Section \ref{sec:TCEx} provides the theoretical analysis of the Toeplitz covariance matrix with ones along the diagonal and correlation coefficient $\rho$'s on the off-diagonals.
The model covariance matrix for a given order, $p$, as well as the KL divergence between the original distribution and the model distribution are also presented in this section.
Moreover, asymptotic upper bounds for KL divergence and the AUC are also presented in this section.
In section \ref{sec:sim} we present some simulation results for approximation model with different orders and investigates the quality of different model approximations based on the numerically evaluated AUC and also its analytical upper and lower bounds.
Finally, Section \ref{sec:con} summarizes results of this paper and discuss further research directions.

\noindent{\bf Notation remark: } In the rest of this paper, with abuse of notion, when we use the KL divergence between random vectors it means the KL divergence between their associated distributions.

\section{Detection Problem Framework}
\label{sec:prob}

\subsection{Preliminaries}

Let $\uX \sim \mcN (\uzm , \bSigma_{\uX})$, i.e. jointly Gaussian with mean 0 and covariance matrix $\bSigma_{\uX}$, where $\uX \in \mbbR^n$.
We want to approximate the random vector $\uX$, with another random vector, $\uXM \in \mbbR^n$ which has a zero-mean jointly Gaussian distribution with the covariance matrix $\bSigma_{\uXM}$ associated with the desired model\footnote{Examples of possible models: star structure and Markov chain.}, i.e. $\uX \sim \mcN (\uzm , \bSigma_{\uXM})$. Note that the model covariance matrix is also positive-definite, $\bSigma_{\uXM} \succ 0$.
Also, let $\mcG=(\mcV, \mcE_\mcM)$ be the graph representation of the model random vector $\uXM$ where sets $\mcV$ and $\mcE_\mcM \subseteq \psi$ are the set of all vertices and the set of all edges of the graph representing of $\uXM$, respectively where $\psi$ is the set of all edges in complete graph with vertex set $\mcV$.

We define the correlation approximation matrix (CAM) associated with the model selection problem as follows.
\begin{Definition}{\bf Correlation approximation matrix \cite{MQ16ArXiv}.}
The CAM for the model is defined as $\bDelta \triangleq \bSigma_{\uX} \bSigma_{\uX_{\mcM}}^{-1}$. \hfill \ensuremath{\blacksquare}
\end{Definition}


\noindent {\bf Remark:} The CAM is a positive definite matrix and its eigenvalues contains all information necessary to compute cost functions associated with the model selection problem.

\subsection{General Framework}

A common measure to compare two probability distributions is the KL divergence. Here we expand the comparison by considering a detection problem where the null hypothesis represents the original random vector, $\uX$ and the alternate hypothesis represents the approximate random vector $\uXM$.
%
%
We need to define a test statistic to quantify the detection problem.
The likelihood ratio test (the Neyman-Pearson (NP) Lemma \cite{np1928}) is the most powerful test statistic where we first define the log-likelihood ratio test (LLRT) as
\begin{equation*}
l(\ux) = \textnormal{log } \frac{f_{\uX}(\ux|\mcH_1)}{f_{\uX}(\ux|\mcH_0)}
\end{equation*}
where $f_{\uX}(\ux|\mcH_0)$ is the distribution of random vector $\uX$ under the null hypothesis while $f_{\uX}(\ux|\mcH_1)$ is the distribution of random vector $\uX$ under the alternative hypothesis.
Moreover, let $L(\uX)$ be the LLRT random variable. Also, let random variables $$L_0 \triangleq L(\uX) | \mcH_0$$ and $$L_1 \triangleq L(\uX) | \mcH_1$$ be the LLRT statistics under hypothesis $\mcH_0$ and hypothesis $\mcH_1$, respectively.
We then define {\it the false-alarm probability} and {\it the detection probability} by comparing the LLRT statistic under each hypothesis with a given threshold, $\uptau$,  and computing the following probabilities
\begin{itemize}
\item[-] The false-alarm probability, $P_{0}(\uptau)$, under the null hypothesis, $\mcH_0$: $P_{0}(\uptau) = \tPr ( L_0 \geq \uptau )$, 
\item[-] The detection probability, $P_{1}(\uptau)$, under the alternative hypothesis, $\mcH_1$: $P_{1}(\uptau) = \tPr ( L_1 \geq \uptau )$. 
\end{itemize}
The most powerful test is defined by setting the false-alarm rate $P_0 (\uptau) = \bar{P_0}$ and then computing the threshold value $\uptau_0$ such that $\Pr(L_0 \geq \uptau_0) = \bar{P_0}$.

\begin{Definition}
The KL divergence between two multivariate continuous distributions $p(\uX)$ and $q(\uX)$ is defined as
$$\mcD \left( p_{\uX}(\ux)||q_{\uX}(\ux) \right) = \int_{\mcX} p_{\uX}(\ux) \log \frac{p_{\uX}(\ux)}{q_{\uX}(\ux)} \; d \ux$$
where $\mcX$ is the feasible set. \hfill \ensuremath{\blacksquare}
\end{Definition}

Throughout this paper we may use other notations such as the KL divergence between two random vectors or the KL divergence between two covariance matrices for zero-mean Gaussian distribution case in order to present the KL divergence between two distributions. 

\begin{Proposition}
Expectation of the LLRT statistic under each hypothesis is
\begin{itemize}
\item[-] $\exc \left(L_0 \right) = - \mcD( f_{\uX}(\ux|\mcH_0) || f_{\uX}(\ux|\mcH_1 ))$,
\item[-] $\exc \left(L_1 \right) = \mcD( f_{\uX}(\ux|\mcH_1) || f_{\uX}(\ux|\mcH_0 ))$.
\end{itemize}
\proof
Proof is based on the KL divergence definition. \hfill \ensuremath{\blacksquare}
\end{Proposition}

The NP decision rule in a regular detection problem framework is to accept the hypothesis $\mcH_1$ if the LLRT statistic, $L(\ux)$, exceeds a critical value which is set based on the false-alarm probability, and reject it otherwise. 
As it is mentioned in \cite{MQ16ArXiv}, we pursue a different goal in the approximation problem scenario. 
We approximate a model distribution, $f_{\uXM}(\ux)$, as close as possible to the given distribution, $f_{\uX}(\ux)$.
In ideal case where there is no approximation error, the detection probability must be equal to the false-alarm probability for the optimal detector at all possible thresholds, i.e. the receiver operating characteristic (ROC) curve \cite{scharfSSP} that represents best detectors for all threshold values should be a line of slope $1$ passing through the origin.

\subsection{Multivariate Gaussian distribution}

Let the random vector $\uX$ have a multivariate Gaussian distribution with covariance matrix $\bSigma_{\uX}$.
In this paper, the null hypothesis, $\mcH_0$, is the hypothesis that the parameter of interest, which is the covariance matrix of the random vector $\uX$, is known and is equal to $\bSigma_{\uX}$ while the alternative hypothesis, $\mcH_1$, is the hypothesis that the random vector $\uX$ is replaced by the model random vector $\uXM$ which means that the random vector $\uX$ has the model approximation distribution with the covariance matrix, $\bSigma_{\uXM}$.
Thus, we can rewrite the LLRT statistic as
\begin{align*}
l(\ux)  = \textnormal{log } \frac{f_{\uXM}(\ux)}{f_{\uX}(\ux)}  
\end{align*}
The LLRT statistic can be simplified for the multivariate Gaussian distributed random vectors as 
\begin{equation}
\label{eq:LLRT}
l(\ux)  = \textnormal{log } \frac{\mcN (\uzm , \bSigma_{\uXM})}{\mcN (\uzm , \bSigma_{\uX})} = - c +  k(\ux)
\end{equation}
where $c = - \frac{1}{2} \textnormal{log } (|\bDelta|)$ is a constant and
$k(\ux) = \ux^T \bK \ux$
where $\bK = \frac{1}{2}( \bSigma_{\uX}^{-1} - \bSigma_{\uXM}^{-1})$ is an indefinite matrix with both positive and negative eigenvalues.

\begin{Theorem}
	\label{thm:covselect}
	{\bf Covariance Selection \cite{dempster}. } Given a multivariate Gaussian distribution with covariance matrix $\bSigma_{\uX}\succ 0$, $f_{\uX}(\ux)$, and a model $\mcM$, there exists a unique approximated multivariate Gaussian distribution with covariance matrix $\bSigma_{\uXM}\succ 0$, $f_{\uXM}(\ux)$, that minimize the KL divergence, $\mcD(f_{\uX}(\ux)||f_{\uXM}(\ux))$ and satisfies the covariance selection rules, i.e. the model covariance matrix satisfies the following covariance selection rules
	\begin{itemize}
		\item[-] $\bSigma_{\uXM}(i,i) = \bSigma_{\uX}(i,i)$, \hspace{0.6cm}$\forall \; i \in \mcV$ 
		\item[-] $\bSigma_{\uXM}(i,j) = \bSigma_{\uX}(i,j)$, \hspace{0.5cm}$\forall \; (i,j) \in \mcE_\mcM$ 
		\item[-] $\bSigma_{\uXM}^{-1}(i,j) = 0$,                 \hspace{1.6cm}$\forall \; (i,j) \in \mcE_\mcM^c$ 
	\end{itemize}
	where the set $\mcE_\mcM^c = \psi - \mcE_\mcM$ represents the complement of the set $\mcE_\mcM$. 
	\proof Proof for Gaussian distributions is given in Dempster 1972 paper \cite{dempster}. \hfill \ensuremath{\blacksquare}
\end{Theorem}

\noindent {\bf Remark:} From theorem \ref{thm:covselect} and definition of the KL divergence for Gaussian distributions, we have $c = \mcD(f_{\uX}(\ux)||f_{\uXM}(\ux))$, since given any covariance matrix and its model covariance matrix satisfying theorem \ref{thm:covselect}, we have $tr(\bDelta) = n$.

\subsection{Distribution of the LLRT statistic}

The random vector $\uX$ has Gaussian distribution under both hypotheses $\mcH_{0}$ and $\mcH_{1}$. 
Thus under both hypotheses, the real random variable, $K(\uX) \triangleq \uX^T \bK \uX$ has a generalized chi-squared distribution, i.e. the random variable, $K(\uX)$, is equal to a weighted sum of chi-squared random variables with both positive and negative weights under both hypotheses.
Let us define $\uW = \bSigma_{\uX}^{-\frac{1}{2}} \uX$ under $\mcH_{0}$ and $\uZ = \bSigma_{\uXM}^{-\frac{1}{2}} \uX$ under $\mcH_{1}$, where $\bSigma_{\uX}^{\frac{1}{2}}$ and $\bSigma_{\uXM}^{\frac{1}{2}}$ are the square root of covariance matrices $\bSigma_{\uX}$ and $\bSigma_{\uXM}$, respectively. 
Then let random vectors $\uW \sim \mcN (\uzm , \bI)$ and  $\uZ \sim \mcN (\uzm , \bI)$ have zero-mean Gaussian distributions with the same covariance matrices, $\bI$, where $\bI$ is the identity matrix of dimension $n$. 
Note that, the CAM is a positive definite matrix with $\lambda_i >0$ where $1\leq i \leq n$.
Thus, the random variable $K(\uX)$, under both hypotheses $\mcH_0$ and  $\mcH_1$ can be defined as
$$ K_0 \triangleq \frac{1}{2} \sum_{i=1}^{n} (1 - \lambda_i) W_i^2$$
and
$$ K_1 \triangleq \frac{1}{2} \sum_{i=1}^{n} (\lambda_i^{-1} - 1) Z_i^2$$
respectively, where random variables $W_i$ and $Z_i$, are the $i$-th element of random vectors $\uW$ and $\uZ$, respectively. 
Moreover, random variables $W_i^2$ and $Z_i^2$, follow the first order central chi-squared distribution.
Note that, $ L_0 = - c + K_0 $ and $ L_1 = - c + K_1 $.

\noindent {\bf Remark:} As a simple consequence of the covariance selection theorem, the summation of weights for the generalized chi-squared random variable, the expectation of $K(\uX)$, is zero under the hypothesis $\mcH_0$, i.e. $\exc (K_0)  = \frac{1}{2} \sum_{i=1}^{n} (1 - \lambda_i) = 0$ \cite{dempster}, and this summation is positive under the hypothesis  $\mcH_1$, i.e. $\exc (K_1)  = \frac{1}{2} \sum_{i=1}^{n} (\lambda_i^{-1} - 1) \geq 0$.

\subsection{Area under the curve}

\begin{figure}[ht]
\begin{centering}
\includegraphics[width=.9\linewidth]{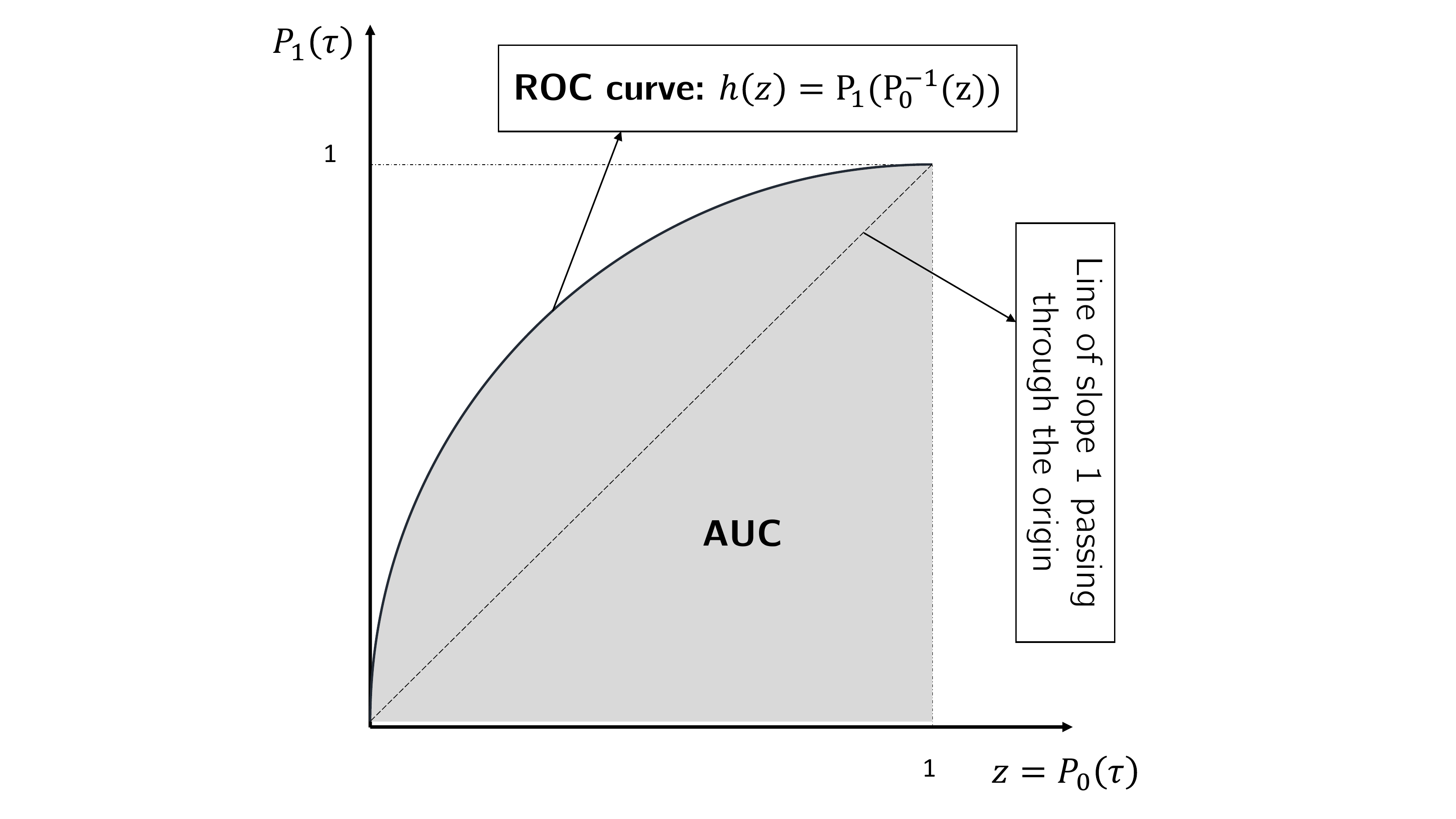}
\caption{The ROC curve and the area under the ROC curve. Each point on the ROC curve indicates a detector with given detection and false-alarm probabilities.}
\label{fig:ROCcurve}
\end{centering}
\end{figure}

As we mentioned before, in approximation set up the desired goal is that the ROC curve is as close as possible to the line of slope $1$ passing through the origin in comparison to the step function in the hypothesis testing problem \cite{MQ16ArXiv}.
Area under the curve is defined as the integral of the ROC curve. Note that in approximation problem presented here we want it to be around half.
Area under the curve (AUC) is defined as the integral of the ROC curve (figure \ref{fig:ROCcurve}) and is a measure of accuracy in decision problems.

\begin{Definition}
The area under the ROC cure (AUC) is defined as 
\begin{equation}
\label{eq:AUCdef}
AUC = \int_{0}^{1} h(z)\,  d\, z = \int_{0}^{1} P_1(\uptau) \, d P_0(\uptau)
\end{equation}
where $\uptau$ is the detection problem threshold.\hfill \ensuremath{\blacksquare}
\end{Definition}

\noindent {\bf Remark:} The AUC is a measure of accuracy for the detection problem and $\sfrac{1}{2} \le {\rm AUC} \le 1$. Note that, in conventional decision problems, the AUC is desired to be as close as possible to $1$ while in approximation problem presented here we want the AUC to be close to $\sfrac{1}{2}$.

\begin{Theorem}{\bf Statistical property of AUC \cite{hanleyAUC}.}
\label{thm:AUC}
The AUC for the LLRT statistic, $L(\uX)$, and two hypotheses, $\mcH_0$ and $\mcH_1$ is 
\begin{equation*}
AUC = \textnormal{Pr} \left( L_{\Delta} > 0  \right).
\end{equation*}
where $ L_{\Delta} \triangleq L_1 -  L_0$.
\hfill\ensuremath{\blacksquare}
\end{Theorem}

\section{Toeplitz Covariance Matrix}
\label{sec:TCEx}

Here, we assumed that the $n$ by $n$ covariance matrix $ \bSigma_{\uX}$ has a Toeplitz structure with ones on the diagonal and the correlation coefficient $\rho$ as off diagonal elements
\begin{equation*}
 \bSigma_{\uX} =
\begin{bmatrix}
                1 &  \rho & \ldots       &\rho      \\
                \rho  & \ddots  & \ddots  &  \vdots          \\
              \vdots & \ddots &   \ddots  & \rho \\
         \rho  &   \ldots   &    \rho   &  1 \\
\end{bmatrix}
.
\end{equation*}

\begin{Definition}
{\bf Clique.} A maximal subset of the nodes which defines a complete subgraph is the clique subgraph.
\hfill \ensuremath{\blacksquare}
\end{Definition}
In other words, all pairs of nodes are connected in the clique subgraph.

\begin{Definition}
{\bf Junction tree.} A junction tree is a clique tree \cite{CliqueTree93} such that for each pair of cliques $C_1$ and $C_2$ in the graph, all cliques on the path between $C_1$ and $C_2$ contain their intersection, $C_1 \cap C_2$.
\hfill \ensuremath{\blacksquare}
\end{Definition}

In this example, we are interested in models which can be represented using junction trees whose vertices are cliques of the size at most $p$.
\footnote{We avoid cycles by turning subsets of the nodes into supernodes.}
Going back to the model selection problem for the example, we are investigating the following two generalizations of the chain and the star networks. 
Note that, we can cunstruct a junction tree for these two special models.

\subsection{$p$th order star network}
The model covariance matrix for the $p$th order star network where all nodes are connected to the first $p$ nodes which all are connected together is as follow
\begin{equation*}
 \bSigma_{\uX_\mcM}^{pth-star} =
\begin{bmatrix}
             1        & \rho  & \ldots & \ldots & \ldots & \ldots &  \rho     \\
             \rho    & \ddots  & \ddots &  &  &  &  \vdots     \\
             \vdots & \ddots  & 1 & \rho & \ldots & \ldots &  \rho     \\
             \vdots & & \rho    & 1  &         \rho_{{}_{1}}       &  \ldots  &   \rho_{{}_{1}}     \\
             \vdots & & \vdots &  \rho_{{}_{1}}    &  \ddots     &  \ddots&  \vdots \\
             \vdots & & \vdots & \vdots   &  \ddots     & \ddots & \rho_{{}_{1}}    \\ 
             \rho    & \ldots & \rho    &  \rho_{{}_{1}}   &  \ldots &  \rho_{{}_{1}}  &  1  \\
\end{bmatrix}
\end{equation*}
where 
$$\rho_{{}_{1}}  =  \frac{p \rho^2 }{(p-1) \rho +1 }.$$

\subsection{$p$th order Markov chain network}
The model covariance matrix for the $p$th order Markov chain network is as follow
\begin{equation*}
 \bSigma_{\uX_\mcM}^{pth-chain} =
\begin{bmatrix}
             1        & \rho  & \ldots &\rho & \rho_{{}_{1}} & \ldots &  \rho_{{}_{n-p-1}}   \\
             \rho    & \ddots  & \ddots &  & \ddots & \ddots &  \vdots     \\
             \vdots & \ddots  & \ddots & \ddots &  & \ddots &  \rho_{{}_{1}}      \\
             \rho & & \ddots    & \ddots  &         \ddots       &    &   \rho   \\
             \rho_{{}_{1}}   & \ddots&  &  \ddots    &  \ddots     &  \ddots&  \vdots \\
             \vdots & & \ddots &    &  \ddots     & \ddots & \rho  \\ 
             \rho_{{}_{n-p-1}}     & \ldots & \rho_{{}_{1}}      &  \rho   &  \ldots &  \rho &  1  \\
\end{bmatrix}
.
\end{equation*}
To satisfy Theorem \ref{thm:covselect}  we have that $\rho_{{}_{i}}$ for $ i \in \{1 , \ldots , n-p-1 \}$ can be computed through the following recursive equation
\begin{equation}
\label{eq:recEQ}
\rho_{{}_{i}} =\urho_{{}_{i-1}}^T \, \uv_i \, \frac{\rho}{(p-1) \rho +1 } 
\end{equation}
where $\uv_i = [ \overbrace{1 , \ldots , 1}^{p} , 0, \ldots, 0]^T$ is a vector of length $n$ and $\urho_{{}_{i}} = [ \rho_{{}_{i}}, \ldots, \rho_{{}_{1}}, \overbrace{\rho , \ldots , \rho}^{p} ]^T $ where $\urho_{{}_{0}} = [ \overbrace{\rho , \ldots , \rho}^{p} ]^T $ is the initialization step.


\begin{Lemma}
\label{lem:pthKL}
The KL divergence for the $p$th order star network and the $p$th order Markov chain network can be calculated as
\begin{align*}
\mcD (\uX||\uX_{pth-chain}) & = \frac{1}{2} (n-p) \log \left(\frac{p \rho + 1 }{(p-1) \rho + 1 } \right) \\
& +  \frac{1}{2} \log \left(\frac{(p-1) \rho + 1 }{(n-1) \rho + 1 } \right) 
\end{align*}
and 
$$\mcD (\uX||\uX_{pth-star}) = \mcD (\uX||\uX_{pth-star}). $$
\proof
Note that, from \cite{Lbanded_kavcic} we have
\begin{equation*}
|  \bSigma_{\uX_\mcM}^{pth-chain} | = \frac{\left[ (p \rho + 1) (\rho - 1)^p \right]^{(n-p)}}{\left[ ( (p-1) \rho + 1) (\rho - 1)^{p-1} \right]^{(n-p-1)}}
\end{equation*}
and
\begin{equation*}
|  \bSigma_{\uX} | =  ( (n-1) \rho + 1) (\rho - 1)^{n-1}.
\end{equation*}
Inserting the values of these determinants into the KL divergence
$$\mcD (\uX||\uX_{\mcM}) = - \frac{1}{2} \log \left( \bSigma_{\uX} \bSigma_{\uX_\mcM}^{-1} \right)$$
we conclude the result for the $p$th order Markov chain network.
To show that the KL divergence for the $p$th order star network is exactly equal to the KL divergence for the $p$th order chain network, we need to construct the corresponding junction tree for each of these networks by grouping appropriate $p$ nodes. Note that, the KL divergence for the junction trees are equal since the mutual information between the junction nodes are exactly equal. 
\hfill \ensuremath{\blacksquare}
\end{Lemma}

\begin{Proposition}
\label{prop:KLbounded}
The KL divergence for the $p$th order star network and the $p$th order Markov chain network is bounded as $n$ goes to infinity if for a given constant number, $\kappa>1$, the order, $p$, is the integer number in interval, 
$$\mcD (\uX||\uX_{pth-star}) < \infty \quad\quad \textnormal{as} \quad\quad  (n \rightarrow \infty, \;\sfrac{n}{p} \rightarrow \kappa ).$$
\proof
Let $p = \ceil[\big]{\sfrac{n}{\kappa}}$ be the smallest integer greater than or equal to $\sfrac{n}{\kappa}$.
The KL divergence can be bounded as follow
\begin{align*}
\mcD (\uX||\uX_{pth-star}) & = \frac{(n-\ceil[\big]{\sfrac{n}{\kappa}}) }{2} \log \left(1+\frac{ \rho }{(\ceil[\big]{\sfrac{n}{\kappa}}-1) \rho + 1 } \right) \\
& +  \frac{1}{2} \log \left(\frac{(\ceil[\big]{\sfrac{n}{\kappa}}-1) \rho + 1 }{(n-1) \rho + 1 } \right) \\
& \stackrel{(a)}{\leq} \frac{(n-\sfrac{n}{\kappa}) }{2} \log \left(1+\frac{\rho }{(\sfrac{n}{\kappa}-1) \rho + 1 } \right) \\
& +  \frac{1}{2} \log \left(\frac{((\sfrac{n}{\kappa}+1)-1) \rho + 1 }{(n-1) \rho + 1 } \right)\\
& \stackrel{(b)}{\leq} \frac{(1-\sfrac{1}{\kappa}) n }{2} \left(\frac{\rho}{(\sfrac{n}{\kappa}-1) \rho + 1 } \right) \\
& +  \frac{1}{2} \log \left(\frac{(\sfrac{n}{\kappa}) \rho + 1 }{(n-1) \rho + 1 } \right)
\end{align*}
Where (a) is true since for the integer order, $p$, we have $\sfrac{n}{\kappa} \leq p < \sfrac{n}{\kappa}+1$
and (b) is true since
$\log (1+z) \leq z $ for $z \geq 0$.
Then, in the limit we have
\begin{align*}
\lim_{n \rightarrow \infty} \mcD (\uX||\uX_{pth-star}) & \leq \frac{(1-\sfrac{1}{\kappa}) }{\sfrac{2}{\kappa}} + \frac{1}{2} \log \left( \sfrac{1}{\kappa} \right) \\
& \leq \frac{\kappa-1}{2} -  \frac{\log (\kappa)}{2}  < \infty
\end{align*}
which complete the proof.\hfill \ensuremath{\blacksquare}
\end{Proposition}

\begin{Proposition}
The AUC of the $p$th order star network and the $p$th order Markov chain network is bounded from $1$ as $n$ goes to infinity if $n = \kappa p$ and $p = \ceil[\big]{\sfrac{n}{\kappa}}$,
$$\textnormal{Pr} \left( L_{\Delta}>0 \right) < 1 . $$
\proof
We can conclude this result from the proposition \ref{prop:KLbounded} upper bound for the KL divergence combined with the upper bound for the AUC,
\begin{align*}
\textnormal{Pr} \left( L_{\Delta}>0 \right) & \leq 1 - e^{- \lim_{n \rightarrow \infty}  \mcD (\uX||\uX_{pth-star})-1} \\
& < 1
\end{align*}
provided in \cite{MQ16ArXiv}.\hfill \ensuremath{\blacksquare}
\end{Proposition}


\section{Simulation Results and Discussion}
\label{sec:sim}

In this section, we consider the Toeplitz example presented before as the covariance matrix for a Gaussian random vector. We calculate different models such as the $p$th order Markov chain and the $p$th order star networks for various values of $p$. 
For a given order, both of the aforementioned models have the same KL divergence values as calculated in lemma \ref{lem:pthKL}.
Moreover, we compute AUC and compare it with its lower and upper bounds \cite{MQ16ArXiv} for these cases.

\begin{figure}[ht]
\begin{minipage}[b]{0.48\linewidth}
\includegraphics[width=1.1\linewidth]{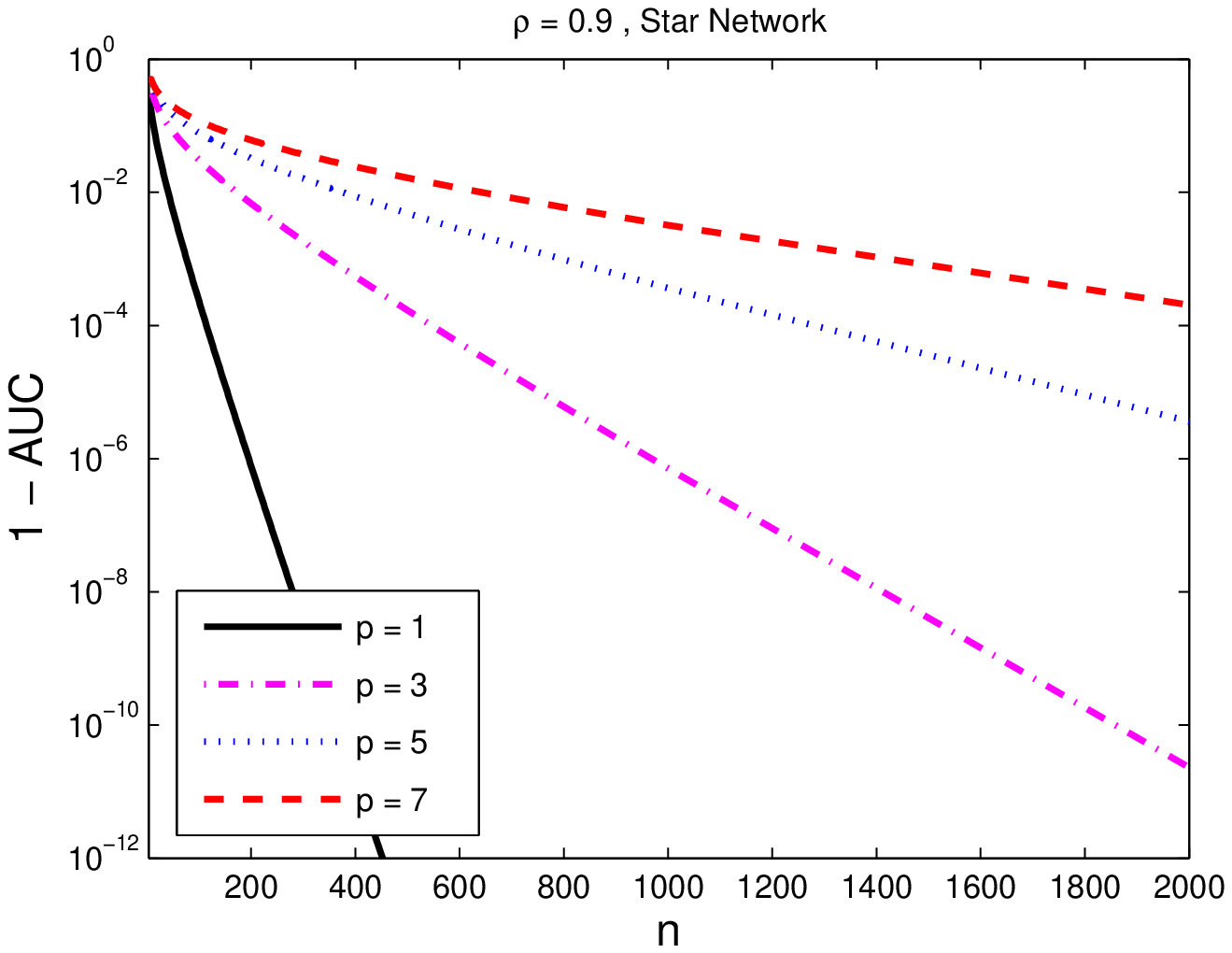}
\end{minipage}
\hspace{0.1cm}
\begin{minipage}[b]{0.48\linewidth}
\includegraphics[width=1.1\linewidth]{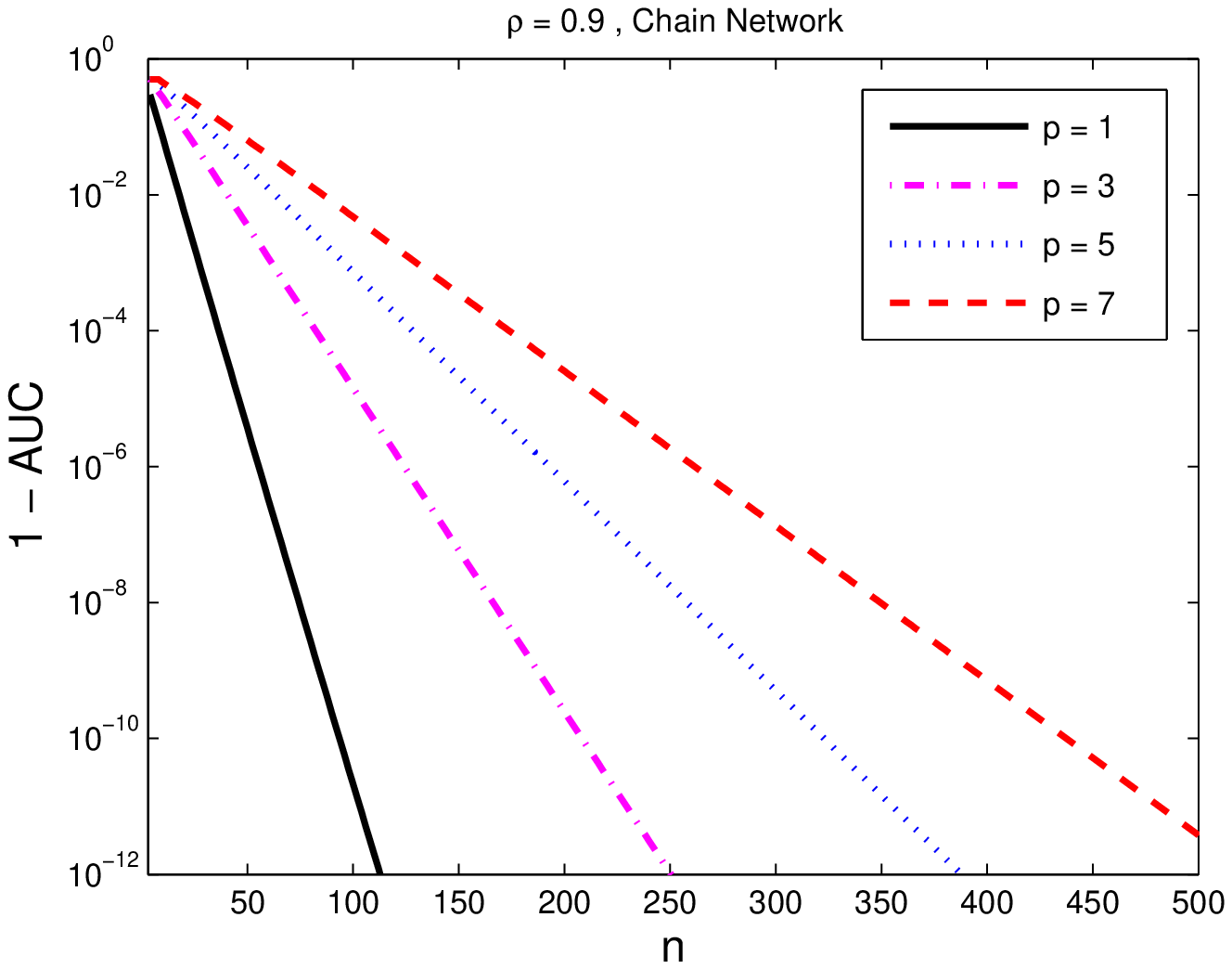}
\end{minipage}
\caption{1 - AUC (log-scale) v.s. the dimension of the graph (linear-scale), $n$, for star approximation {\bf (left)} and chain approximation {\bf (right)} with different model orders, $p=1$, $p=3$, $p=5$ and $p=7$ and correlation coefficient $\rho = 0.9$.}
\label{fig:Toeplitz_Example_comp_ord_Linear}
\end{figure}

Figure \ref{fig:Toeplitz_Example_comp_ord_Linear} plots (1 - AUC) in log-scale v.s. the dimension of the graph, $n$, in linear-scale for star approximation {\bf (left)} and chain approximation {\bf (right)} with different model orders, $p=1$, $p=3$, $p=5$ and $p=7$ for correlation coefficient $\rho = 0.9$. As it is indicated in this figure, (1 - AUC) decreases as the order of the model increases for both star and chain models. Moreover, from this figure, we can conclude that the $p$th order star network performs better than the $p$th order Markov chain network since (1 - AUC) decay exponent is smaller for the former model than the latter model. This can also be seen by comparing the covariance matrix $\bSigma_{\uX}$ and the model covariance matrix, $\bSigma_{\uXM}$ where the model covariance matrix associated with the $p$th order star network is more similar to the covariance matrix $\bSigma_{\uX}$ than the model covariance matrix associated with the $p$th order Markov chain network.
For example, even the quality of the first order star network approximation is better than the quality of the fifth order Markov chain approximation in the simulation results provided in this figure.

\begin{figure}[ht]
\begin{center}
\includegraphics[width=\linewidth]{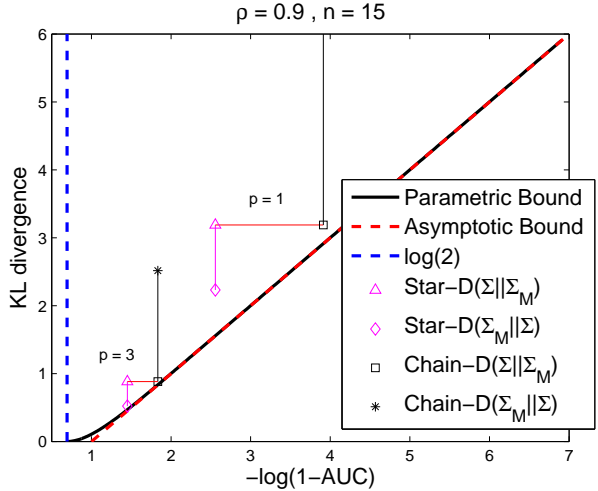}
\caption{KL divergence v.s. AUC and the AUC parametric bound \cite{MQ16ArXiv} v.s. for graph dimension, $n=15$ for the $p$th order Markov chain approximation and $p$th order star network for $p=1$ and $p=3$ with $\rho = 0.9$.}
\label{fig:FeasibleRegion_linear_rho9_p1p3_n15_step}
\end{center}
\end{figure}

Figure \ref{fig:FeasibleRegion_linear_rho9_p1p3_n15_step} plots KL divergence v.s $-\log$ (1 - AUC) for the presented models. In this figure, the dimension $n$ is set to $15$, the order $p$ is set to $1$ and $3$ and the correlation coefficient $\rho$ is set to $0.9$. Furthermore, the feasible region presented in \cite{MQ16ArXiv} and its asymptotic behavior are also plotted in this figure. 
For both models, the KL divergence and the reverse KL divergence are computed and are plotted on this figure. Note that, KL divergences for both models are equal (see lemma \ref{lem:pthKL}) and are connected in this figure. 
As it is shown in the figure, the third order model has better performance than the first order model.

\begin{figure}[ht]
\begin{minipage}[b]{0.48\linewidth}
\includegraphics[width=1.1\linewidth]{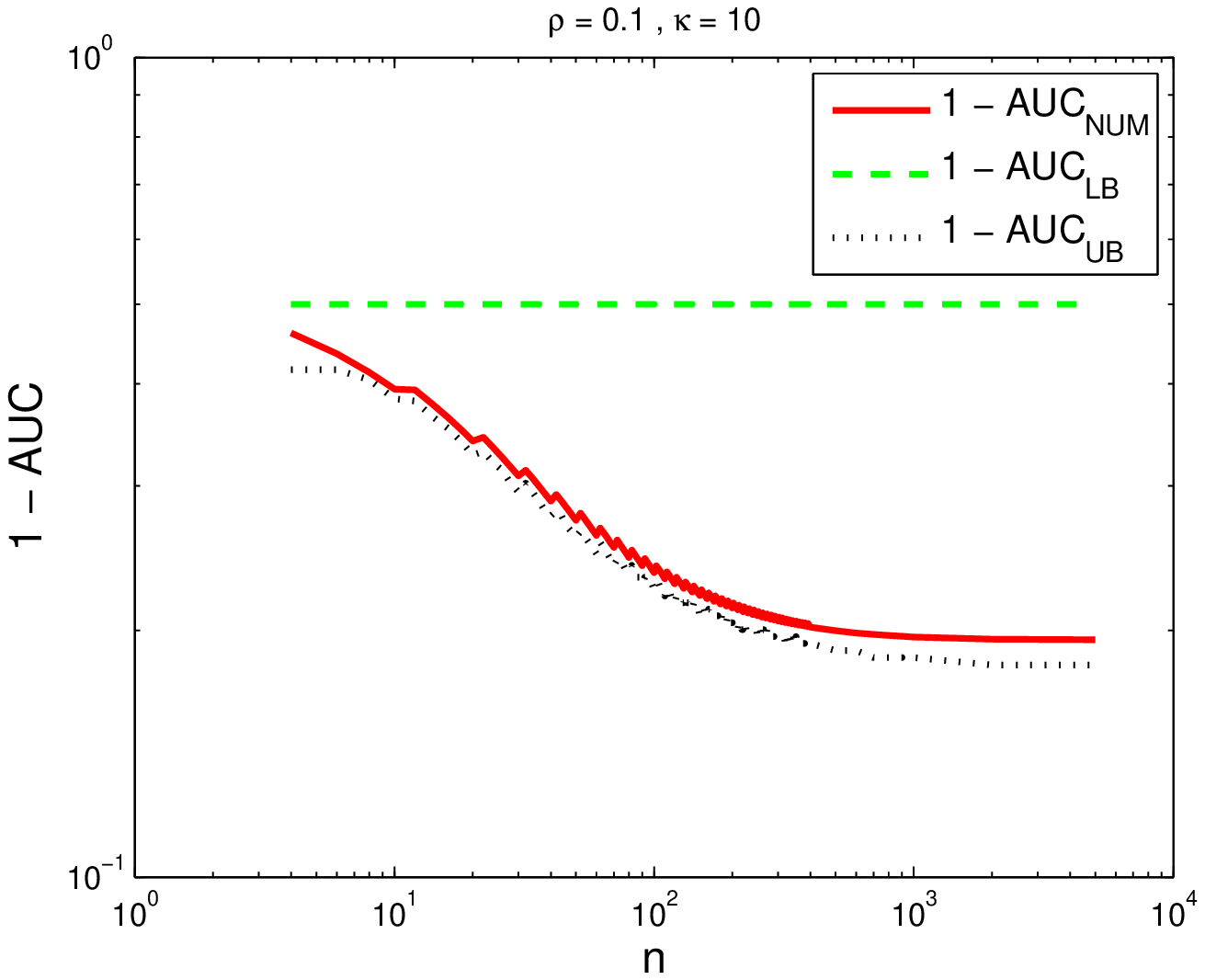}
\end{minipage}
\hspace{0.1cm}
\begin{minipage}[b]{0.48\linewidth}
\includegraphics[width=1.1\linewidth]{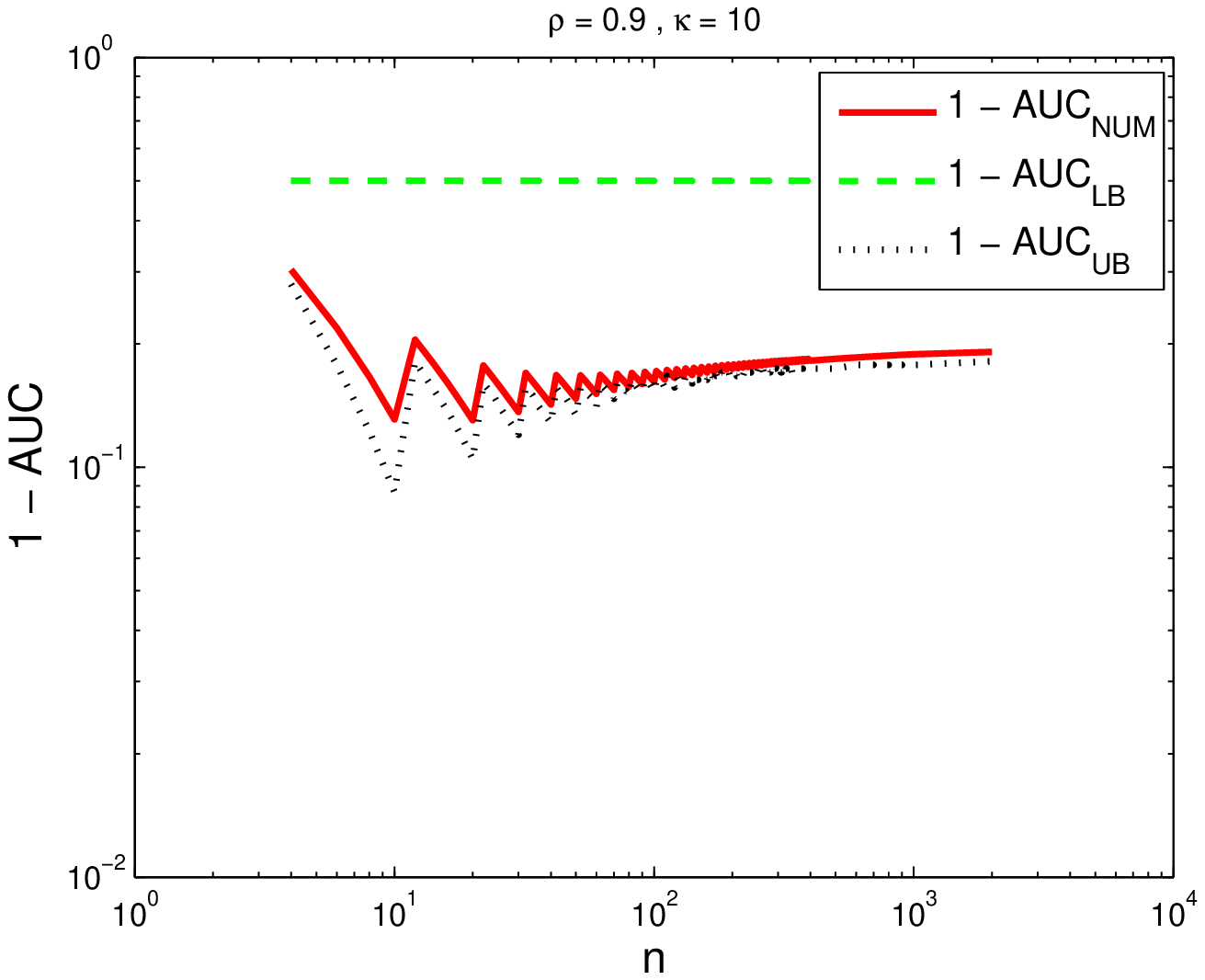}
\end{minipage}
\caption{1 - AUC and its lower and upper bounds v.s. the dimension of the graph, $n$ for the $p$th order star approximation of the Toeplitz example for $\rho = 0.1$ {\bf (left)} and $\rho=0.9$ {\bf (right)} with the model order $p = \ceil[\big]{\sfrac{n}{\kappa}}$ where $\kappa=10$.}
\label{fig:Toeplitz_Example_star_prop_ord}
\end{figure}

\begin{figure}[ht]
\begin{minipage}[b]{0.48\linewidth}
\includegraphics[width=1.1\linewidth]{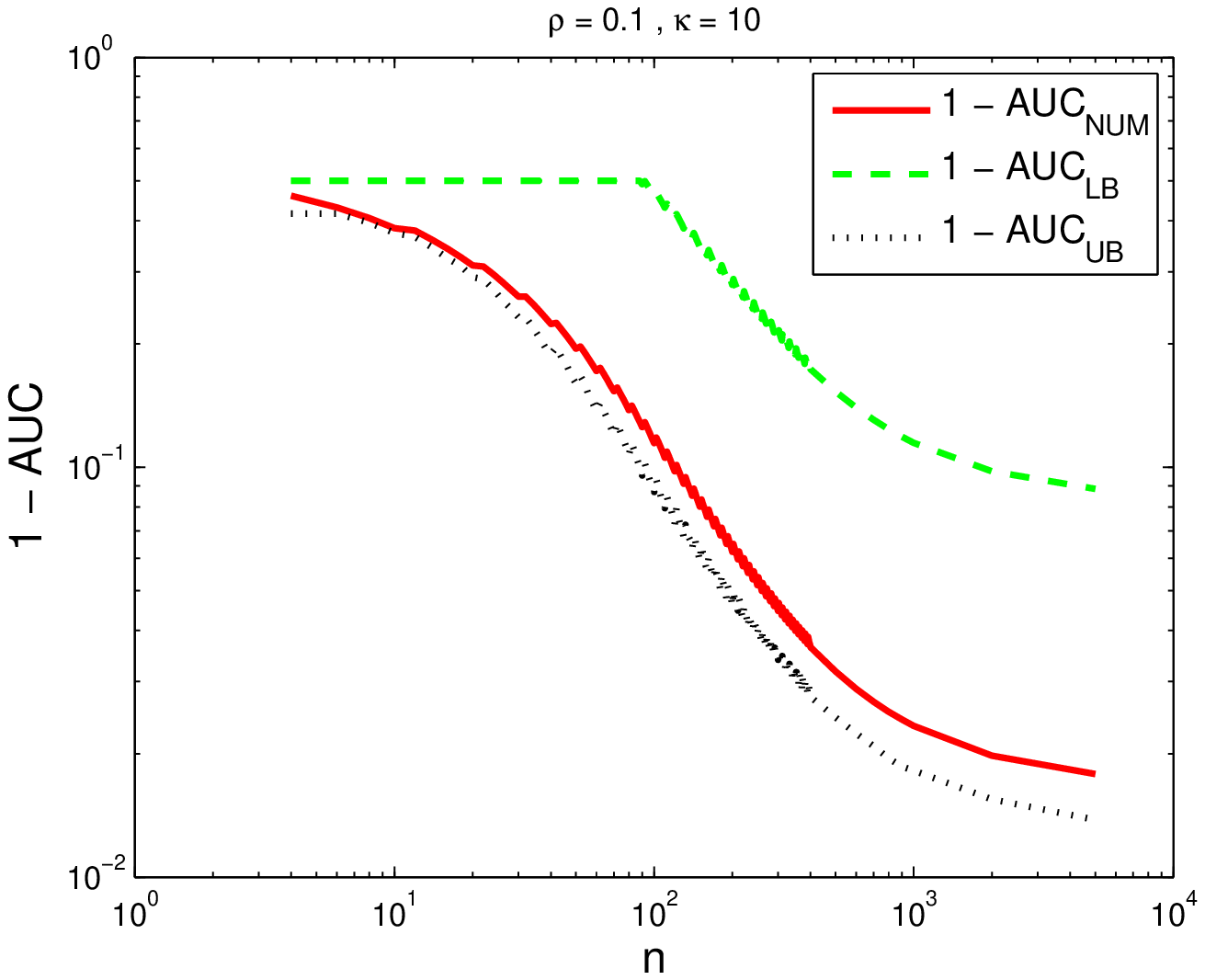}
\end{minipage}
\hspace{0.1cm}
\begin{minipage}[b]{0.48\linewidth}
\includegraphics[width=1.1\linewidth]{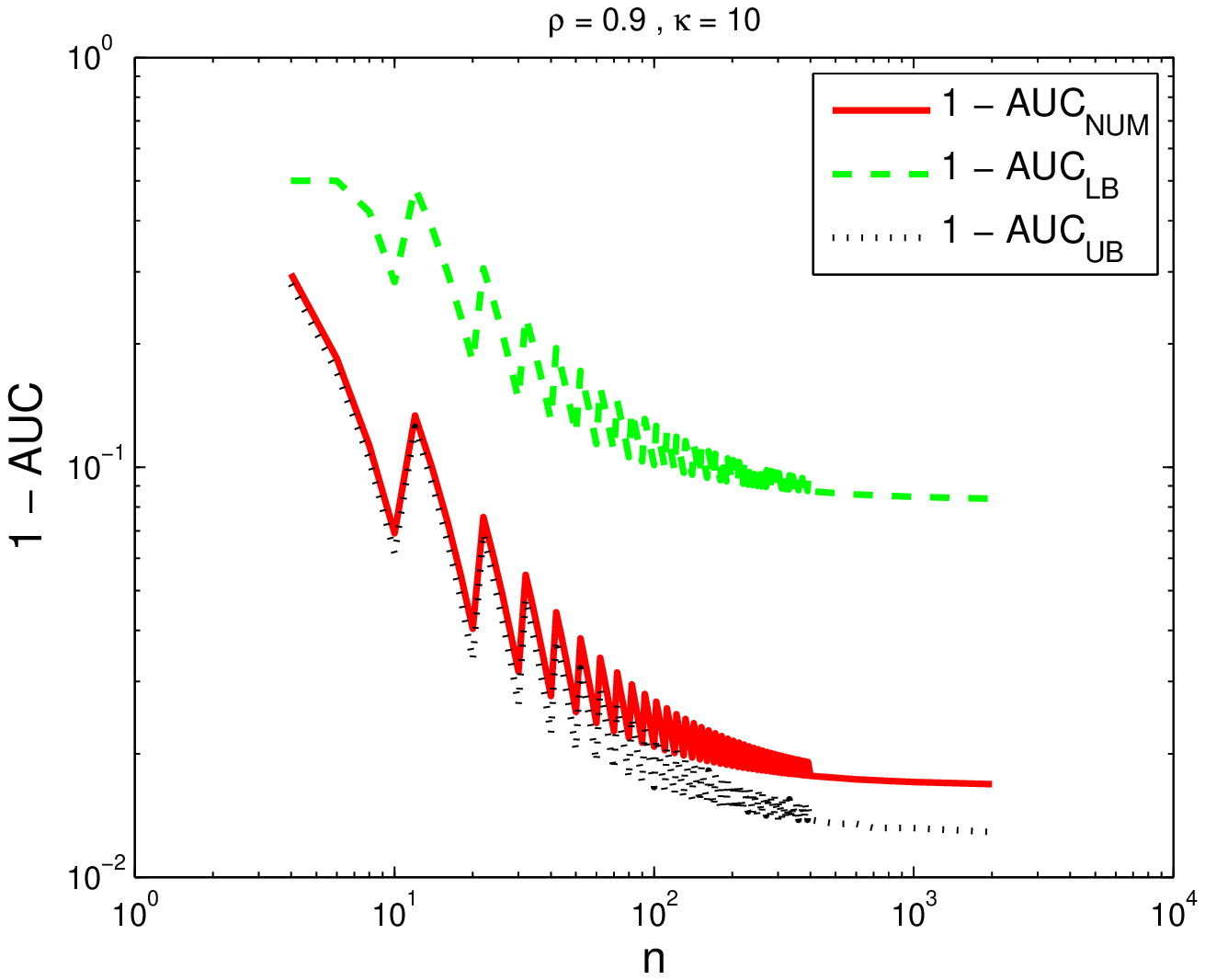}
\end{minipage}
\caption{1 - AUC and its lower and upper bounds v.s. the dimension of the graph, $n$ for the $p$th order Markov chain approximation of the Toeplitz example for $\rho = 0.1$ {\bf (left)} and $\rho=0.9$ {\bf (right)} with the model order $p = \ceil[\big]{\sfrac{n}{\kappa}}$ where $\kappa=10$.}
\label{fig:Toeplitz_Example_chain_prop_ord}
\end{figure}

Figure \ref{fig:Toeplitz_Example_star_prop_ord} plots 1 - AUC v.s. the dimension of the graph, $n$ for the $p$th order star approximation of the Toeplitz example for $\rho = 0.1$ {\bf (left)} and $\rho=0.9$ {\bf (right)} while keeping the model order proportional to the number of nodes in the graphical model, $n$. More specifically, in this figure, we set the model order $p = \ceil[\big]{\sfrac{n}{\kappa}}$ where $\kappa=10$. Moreover, this figure plots the lower bound and the upper bound for 1 - AUC\footnote{Bounds are presented in \cite{MQ16ArXiv}.}. From this figure, we conclude that, $p$th order star approximation is a good approximation model when the model order, $p$ is proportional to the number of nodes, $n$, since the AUC is bounded from one as $n \rightarrow \infty$.
Similarly, figure \ref{fig:Toeplitz_Example_chain_prop_ord} plots 1 - AUC and its upper and lower bounds v.s. the dimension of the graph, $n$ for the $p$th order Markov chain approximation of the Toeplitz example for $\rho = 0.1$ {\bf (left)} and $\rho=0.9$ {\bf (right)} with $p = \ceil[\big]{\sfrac{n}{\kappa}}$ where $\kappa=10$. 
Plots in this figure are not monotonicly decreasing since both the order $p$ and the dimension $n$ are integers and thus the ratio $\sfrac{n}{p}$ is not exactly equal to $\kappa$ for all values of $p$ and $n$.
Furthermore, from the figure, the $p$th order Markov chain approximation is a good approximation model when the model order, $p$ is proportional to the number of nodes, $n$, since the AUC is bounded from one as $n \rightarrow \infty$.
Comparing the plots in figure \ref{fig:Toeplitz_Example_star_prop_ord} and figure \ref{fig:Toeplitz_Example_chain_prop_ord} we can clearly see that even though the AUC for both approximation models are bounded from one, the $p$th order star approximation model is a better model than the $p$th order Markov chain approximation model.

\section{conclusion}
\label{sec:con}


In this paper, we formulate a detection problem to investigate the quality of the graphical model approximation.
We discuss the quality of model selection approximation by examining the area under the curve (AUC).
We consider jointly Gaussian random vectors and give the definition for the correlation approximation matrix (CAM).
We discuss graphical models with junction trees such as the $p$th order Markov chain and the corresponding star network interpretation for a special Toeplitz covariance matrix with ones along the diagonal and correlation coefficient $\rho$'s on the off-diagonals.
These models has very short loops and has associated junction tree that connects cliques of the same size.
The model covariance matrix as well as the KL divergence between the original distribution and the model distribution are computed for the presented Toeplitz covariance matrix.
We also quantify the goodness of the covariance selection problem for this Toeplitz covariance matrix.
For this covariance matrix, we show that if the model order, $p$, is proportional to the number of nodes, $n$, then the model selection is asymptotically good as $ n \rightarrow \infty$ since the AUC is asymptotically bounded away from one.
We conduct some simulations which show that the selected model quality increases as the model order, $p$, increases and confirm our theoretical results.

\section*{Acknowledgment}

\textcolor{black}{This work was supported in part by NSF grant ECCS-1310634, and the University of Hawaii REIS project.}

\bibliographystyle{IEEEbib}
\bibliography{refs}

\end{document}